\documentclass[prd,preprint,tightenlines,floatfix,showpacs,preprintnumbers,nofootinbib,eqsecnum]{revtex4}
 \usepackage[dvips,final]{graphicx}
  \usepackage{amssymb}
   \usepackage{amsmath}
    \usepackage{epsfig}
     \usepackage{bm}

\newcommand{\LI}[1]{\textrm{Li}_{#1}}
\newcommand{\Tr}{\textrm{Tr}}

\begin{document}
\thispagestyle{empty} \preprint{\hbox{}} \vspace*{-10mm}

\title{Scalar Higgs boson production \\ in a fusion of two off-shell gluons}
\author{R.~S.~Pasechnik}
\email{rpasech@theor.jinr.ru}
\author{O.~V.~Teryaev}
\email{teryaev@theor.jinr.ru}

\affiliation{ Bogoliubov Laboratory of Theoretical Physics, JINR,
Dubna 141980, Russia} \affiliation{ Faculty of Physics, Moscow
State University, Moscow 119992, Russia}

\author{A.~Szczurek}
\email{antoni.szczurek@ifj.edu.pl}

\affiliation{Institute of Nuclear Physics PAN, PL-31-342 Cracow,
Poland} \affiliation{University of Rzesz\'ow, PL-35-959 Rzesz\'ow,
Poland}

\date{\today}

\begin{abstract}
The amplitude for scalar Higgs boson production in a fusion of two
off-shell gluons is calculated including finite (not infinite)
masses of quarks in the triangle loop. In comparison to the
effective Lagrangian approach, we have found a new term in the
amplitude. The matrix element found can be used in the
$k_{\perp}$-factorization approach to the Higgs boson production.
The results are compared with the calculations for on-shell
gluons. Small deviations from the $\cos^2 \phi$-dependence are
predicted. The off-shell effects found are practically negligible.
\end{abstract}



\maketitle

\section{Introduction}

The Higgs boson is the only missing, undiscovered component of the Standard
Model of Particle Physics. Within the context of
the Standard Model, the Higgs boson is responsible for the
breaking of the $SU_{L}(2)\times U(1)$ gauge symmetry and provides
the mechanism for the generation of masses of the corresponding
gauge bosons: $W^{\pm}$ and $Z.$ In addition, the same
mechanism provides masses for the leptons and quarks via Yukawa
couplings. Therefore, the discovery and subsequent study of the
Higgs boson properties is of the highest priority for
particles physics community.

A precise theoretical understanding of Higgs production rate
is critical to any attempts to search for the particle.
The dominant production mechanism for Higgs bosons in
hadron-hadron colliders is via gluon-gluon fusion \cite{Georgi}
$pp\rightarrow gg\rightarrow H,$ in which gluons fuse through a
virtual top quark triangle to produce the Higgs. Such a process
provides the largest production rate for the entire Higgs mass range
of interest.

The standard description of hard processes in hadron collisions within the
framework of QCD parton model, which reduces the hadron-hadron
interactions to the parton-parton ones via the formalism of the
hadron structure functions. The most popular approach is the QCD
collinear approximation \cite{collinear}, based on the well known
collinear factorization theorem \cite{Collins1}. In this approach
all particles involved are assumed to be on the mass shell,
carrying only longitudinal momenta, and the cross section is
averaged over two transverse polarizations of the incident gluons.
The transverse momenta of the incident partons are neglected in
the QCD matrix elements. However, at small $x$, the effects of
finite transverse momenta of partons become increasingly
important, especially in the analysis of jets and heavy-quark
production, and there is no reason to neglect the transverse
momenta of the gluons in comparison to the quark mass. The method
to incorporate the incident parton transverse momenta is referred
to as $k_{\perp}$-factorization approach
\cite{CataniCollins,Gribov_Levin}. Here the Feynman diagrams are
calculated taking into account the virtualities and all possible
polarizations of the incident partons. There are widely discussed
applications of the $k_{\perp}$-factorization approach to
hard QCD processes like the $J/\psi$ hadroproduction
\cite{HKSST1}, charmonium production \cite{HKSST2}, heavy
quark photo- \cite{Mariotto,LS04} and hadroproduction \cite{LS05},
Higgs boson production \cite{Lipatov,Szczurek}. Some exclusive
processes in the framework of the $k_{\perp}$-factorization approach are
described in detail in \cite{Nagashima}.

\begin{figure}[h]
 \centerline{\includegraphics[width=0.35\textwidth]{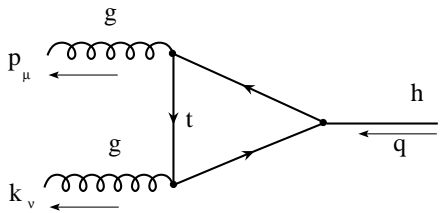}}
   \caption{\label{fig:fig1}
   \small Gluon coupling to the Higgs boson via $t$-quark triangle loop.}
\end{figure}

In the lowest order the gluon coupling to the Higgs boson
in the Standard Model is
mediated by triangular loops with top quarks as shown by the
Feynman diagram in Fig.~\ref{fig:fig1}.
This process for on-shell gluons $p^2,k^2=0$ is well-known in the
literature \cite{WRG}, so we will only present the result
\begin{eqnarray}
 T_{\mu \nu}(k,p)|_{k^{2},p^{2}=0}=i\delta^{ab}\frac{\alpha_{s}}{2\pi}\frac{1}{v}
 ([(kp)g_{\mu \nu}-k_{\mu}p_{\nu}]I_{1} + p_{\mu}k_{\nu}I_{2}),
 \label{on-shellAmp}
 \\
 I_{1}=\tau
 \left[1-\frac14(1-\tau)\ln\left(\frac{\sqrt{1-\tau}+1}{\sqrt{1-\tau}-1}\right)^{2}\right]\quad
 \textrm{for}\quad
 \tau=\frac{4m_{top}^{2}}{m_{h}^{2}}>1.
\label{on-shellFF}
\end{eqnarray}
For a large quark mass $\tau\gg1$ the form factor $I_{1}$ can be expanded
in powers of $\tau^{-1}$
\begin{eqnarray}
 I_{1}=\frac23
 \left[1+\frac{7}{30}\frac{1}{\tau}+\frac{2}{21}\frac{1}{\tau^{2}}+O\left(\frac{1}{\tau^3}\right)\right],\qquad
 I_{1}|_{\tau\rightarrow\infty}=\frac23.
\label{series}
\end{eqnarray}
The form factor $I_{2}$ is not relevant in the collinear approximation
because it does not come into the squared matrix element in
the on-shell limit.

In order to use the $k_{\perp}$-factorization approach we will
take into account the non-trivial virtualities of external gluons.
The main technical problem here is to calculate the vector and
second-rank tensor Feynman integrals with $p^2,k^2\not=0.$ But the
symmetric properties of full amplitude together with a usage of
convenient projectors allow to avoid such difficulties and
represent all relevant form factors in terms of scalar three- and
two-point functions plus extra finite subtractions appearing due to
the regularization.

In the present work we shall study in detail the fusion of
off-shell gluons in proton-proton collisions producing a scalar
Higgs boson by the top quark triangular loops with finite (not
infinite) top quark mass. We analytically calculate the exact
amplitude of this process in terms of two relevant form factors.
The amplitude is studied in various kinematical regions.
Furthermore we discuss different fermion mass limits. We find that
the squared matrix element taking into account the non-zeroth
gluon virtualities slightly (by several percents) increases with
the growth of gluon transverse momentum. We also discuss
consequences of the non-zeroth virtualities for scalar boson
production in the $k_{\perp}$-factorization approach for large
transverse momenta.

\section{The amplitude for the Higgs boson
production in a fusion of off-shell gluons}

Let us start from a general tensor representation of the triangle
amplitude shown in Fig.~\ref{fig:fig1}
\begin{eqnarray}
T_{\mu \nu}(k,p)=g_{\mu \nu}F_{1}+k_{\mu}k_{\nu}F_{2}+
p_{\mu}p_{\nu}F_{3}+k_{\mu}p_{\nu}F_{4}+p_{\mu}k_{\nu}F_{5}.
\label{amp}
\end{eqnarray}

Here $F_{j}=F_{j}(q^{2},k^{2},p^{2},m_{f}^{2}),\;j=1,...,5$ are
the Lorentz invariant form factors, $m_{f}$ is the fermion mass in
the loop. The Bose symmetry of the amplitude $T_{\mu
\nu}(k,p)=T_{\nu \mu}(p,k)$ is equivalent to
\begin{eqnarray}
 F_{1}(k,p) &=&  F_{1}(p,k), \quad
 F_{2}(k,p) =  F_{3}(p,k),
 \nonumber
\\
 F_{4}(k,p) &=&  F_{4}(p,k), \quad
 F_{5}(k,p) =  F_{5}(p,k).
 \label{bose}
\end{eqnarray}

The gauge invariance leads to the vector Ward identities
$p^{\mu}T_{\mu \nu}=0,$ $k^{\nu}T_{\mu \nu}=0$ which in terms of
form factors gives
\begin{eqnarray}
 F_{1}+p^2 F_{3}+(kp)F_{4} &=& 0, \quad
 (kp)F_{2}+p^2 F_{5} = 0, \nonumber
\\
 F_{1}+k^2 F_{2}+(kp)F_{4} &=& 0, \quad
 (kp)F_{3}+k^2 F_{5} = 0.
\label{gauge}
\end{eqnarray}
Therefore we have finally only three relations
\begin{eqnarray}
 F_{1}+p^2 F_{3}+(kp)F_{4} = 0 \; , \nonumber
\\
 (kp)F_{2}+p^2 F_{5} = 0,
\label{equat}
\\
 p^{2} F_{3} - k^2 F_{2} = 0 \; . \nonumber
\end{eqnarray}
This reduces the number of independent form factors to two. So
the tensor representation of the amplitude can be expressed in the
following form
\begin{eqnarray}
T_{\mu \nu}(k,p) &=&
i\delta^{ab}\frac{\alpha_{s}}{2\pi}\frac{1}{v}\biggl([(kp)g_{\mu
\nu}-k_{\mu}p_{\nu}]G_{1}+ \nonumber
\\
&+&\left[p_{\mu}k_{\nu}-\frac{p^{2}}{(kp)}k_{\mu}k_{\nu}-\frac{k^{2}}{(kp)}p_{\mu}p_{\nu}+
\frac{k^{2}p^{2}}{(kp)^2}k_{\mu}p_{\nu}\right]G_{2}\biggr).
\label{ampmain}
\end{eqnarray}
where $\alpha_{s}$ is the strong coupling constant,
$v=(G_{F}\sqrt{2})^{-1/2}$ is the electro-weak mass scale, $a,\,b$
are the color indices for two off-shell gluons. This general form
of the amplitude coincides with the standard expression for
on-shell gluons (\ref{on-shellAmp}) with
\begin{eqnarray}
 I_{1}=G_{1}|_{k^{2},p^{2}\rightarrow0},
\quad I_{2}=G_{2}|_{k^{2},p^{2}\rightarrow0} \; .
 \nonumber
\end{eqnarray}

In order to calculate the form factors $G_{1},\;G_{2}$ we
introduce two projectors. The most convenient and simple choice is
one symmetric $P_{\mu \nu}=g_{\mu \nu}$ and one antisymmetric
$Q_{\mu \nu}=p_{\mu}k_{\nu}-p_{\nu}k_{\mu}$ 2-rank tensors.
Corresponding projections are
\begin{eqnarray}
 S_{1} &=& T_{\mu \nu}Q^{\mu \nu}=i\delta^{ab}\frac{\alpha_{s}}{2\pi}\frac{1}{v}\left([k^{2}p^{2}-(kp)^2]G_{1}+
 \left[2k^{2}p^{2}-(kp)^{2}-\frac{p^{4}k^{4}}{(kp)^{2}}\right]G_{2}\right),
 \nonumber
\\
\label{scalars}
\\ \nonumber
 S_{2} &=& T_{\mu \nu}P^{\mu \nu}=i\delta^{ab}\frac{\alpha_{s}}{2\pi}\frac{1}{v}\left(3(kp)G_{1}+
 \left[(kp)-\frac{k^{2}p^{2}}{(kp)}\right]G_{2}\right). \nonumber
\end{eqnarray}
Solving this system of equation with respect to $G_{1},\;G_{2}$ we get
\begin{eqnarray}
G_{1} &=& [i\delta^{ab}]^{-1}\frac{\pi
v}{\alpha_{s}}\frac{(kp)S_{1}+((kp)^2-k^{2}p^{2})S_{2}}{(kp)((kp)^{2}-k^{2}p^{2})},
 \nonumber
\\
\label{G1G2thrS}
\\ \nonumber
G_{2} &=& -[i\delta^{ab}]^{-1}\frac{\pi
v}{\alpha_{s}}\frac{(kp)(3(kp)S_{1}+((kp)^{2}-k^{2}p^{2})S_{2})}{((kp)^{2}-k^{2}p^{2})^{2}}.
\end{eqnarray}

The explicit expression for the amplitude reads
\begin{eqnarray}
 T_{\mu \nu}(k,p) &=& i\delta^{ab}\frac{4\pi\alpha_{s}}{v}m_{f}\int
 \frac{d^{4}r}{i(2\pi)^4}\frac{M_{\mu
     \nu}(k,p)}{[(r-k)^{2}-m_{f}^{2}][r^2-m_{f}^{2}][(r+p)^{2}-m_{f}^{2}]} \; ,
\nonumber
\\
\nonumber
\\
 M_{\mu \nu}(k,p) &=& \Tr[(m_{f}+\not{r}-\not{k})
 \gamma_{\nu}(m+\not{r})\gamma_{\mu}(m_{f}+\not{r}+\not{p})] \; .
\nonumber
\end{eqnarray}

Employing now the dimensional regularization and using the
Passarino-Veltman reduction we get
\begin{eqnarray}
S_{1} &=& i\delta^{ab}\frac{\alpha_{s}}{\pi
v}\,\frac{m^{2}_{f}}{\mu^{4-n}}\;
[C_{0}(kp)(k^{2}+p^{2}+2(kp))-(k^2+(kp))B^{12}_{0}+ \nonumber
\\
&+& (k^2+p^2+2(kp))B^{13}_{0}-(p^2+(kp))B^{23}_{0}],
\\
\nonumber
\\
S_{2} &=& i\delta^{ab}\frac{\alpha_{s}}{\pi
v}\,\frac{m^{2}_{f}}{\mu^{4-n}}\;
[C_{0}(4m^{2}-k^{2}-p^{2}-n(kp))+B^{12}_{0}+ \nonumber
\\
&+& (2-n)B^{13}_{0}+B^{23}_{0}].
\end{eqnarray}
where $n=4-\varepsilon$ is the dimension of the space,
$B^{12},\,B^{13},\,B^{23}$ are the scalar two-point functions
\begin{eqnarray}
B^{12}_{0}(k^{2},m_{f}^{2}) &=& 16\pi^{2}\mu^{4-n}\int
 \frac{d^{n}r}{i(2\pi)^4}\frac{1}{[(r-k)^{2}-m_{f}^{2}][r^2-m_{f}^{2}]}=
\frac{2}{\varepsilon}+\xi-L_{1}, \nonumber
\\
\nonumber
\\
B^{13}_{0}(q^{2},m_{f}^{2}) &=& 16\pi^{2}\mu^{4-n}\int
 \frac{d^{n}r}{i(2\pi)^4}\frac{1}{[(r-k)^{2}-m_{f}^{2}][(r+p)^{2}-m_{f}^{2}]}=
 \frac{2}{\varepsilon}+\xi-L_{2}, \nonumber
\\
\nonumber
\\
B^{23}_{0}(p^{2},m_{f}^{2}) &=& 16\pi^{2}\mu^{4-n}\int
 \frac{d^{n}r}{i(2\pi)^4}\frac{1}{[r^2-m_{f}^{2}][(r+p)^{2}-m_{f}^{2}]}=
 \frac{2}{\varepsilon}+\xi-L_{3}, \nonumber
\end{eqnarray}
where
\begin{eqnarray}
L_{j} = \beta_{j}\ln\frac{\beta_{j}+1}{\beta_{j}-1},\quad
\beta_{1} &=& \sqrt{1-\frac{4m^{2}_{f}}{k^{2}}},\quad \beta_{2}=
\sqrt{1-\frac{4m^{2}_{f}}{q^{2}}},\quad \beta_{3}=
\sqrt{1-\frac{4m^{2}_{f}}{p^{2}}} \nonumber
\\
\xi&=&-\gamma-\ln\pi+2-\ln\frac{m^2_{f}}{\mu^2} \nonumber
\end{eqnarray}
and $C_{0}$ is the convergent scalar three-point function.
Following Ref. \cite{Veretin}
\begin{eqnarray}
C_{0}(q^{2},k^{2},p^{2},m_{f}^{2}) &=& 16\pi^{2}\mu^{4-n}\int
 \frac{d^{n}r}{i(2\pi)^4}\frac{1}{[(r-k)^{2}-m_{f}^{2}][r^2-m_{f}^{2}][(r+p)^{2}-m_{f}^{2}]}=
\nonumber
\\
&=&
\varkappa(k^{2},p^{2},q^{2})+\varkappa(q^{2},k^{2},p^{2})
+\varkappa(p^{2},q^{2},k^{2}) \; .
\nonumber
\end{eqnarray}
Above for simplicity we have introduced the following notations
\begin{eqnarray}
\varkappa(x,y,z)=\frac{1}{\lambda}\left[\LI{2}\left(\frac{t-1}{t-\tau}\right)+
\LI{2}\left(\frac{t-1}{t+\tau}\right)-\LI{2}\left(\frac{t+1}{t-\tau}\right)-
\LI{2}\left(\frac{t+1}{t+\tau}\right)\right] \; . \nonumber
\end{eqnarray}
\begin{eqnarray}
\lambda(x,y,z)=\sqrt{x^2+y^2+z^2-2xy-2yz-2zx},\quad
t=\frac{1}{\lambda}(x-y-z),\; \tau=\sqrt{1-\frac{4m^{2}_{f}}{x}}
\nonumber
\end{eqnarray}
Of course, the projections $S_{1}$ and $S_{2}$ are convergent in
the limit $\varepsilon\rightarrow 0$ and then
\begin{eqnarray}
S_{1} &=& i\delta^{ab}\frac{\alpha_{s}}{\pi v}\,m^{2}_{f}\;
[C_{0}(kp)(k^{2}+p^{2}+2(kp))+((kp)+k^{2})L_{1}- \nonumber
\\
&-& (k^{2}+p^{2}+2(kp))L_{2}+((kp)+p^{2})L_{3}],
\\
\nonumber
\\
S_{2} &=& i\delta^{ab}\frac{\alpha_{s}}{\pi v}\,m^{2}_{f}\;
[C_{0}(4m_{f}^{2}-k^{2}-p^{2}-4(kp))-L_{1}+2L_{2}-L_{3}+2].
\end{eqnarray}
There is a finite subtraction term in $S_{2}$ as
a consequence of regularization. Substituting these expressions
into (\ref{G1G2thrS}) we finally obtain
\begin{eqnarray}
  G_{1} &=&
  \frac{m^{2}_{f}}{(kp)((kp)^{2}-k^{2}p^{2})}\times
  \nonumber
\\
  &\times&
  [(4m^{2}_{f}((kp)^{2}-k^{2}p^{2})-2(kp)((kp)^2-2k^{2}p^{2})+k^{2}p^{2}(k^{2}+p^{2}))C_{0}+
  \nonumber
\\
  &+&
  k^{2}(p^{2}+(kp))L_{1}-(2k^{2}p^{2}+(kp)(k^{2}+p^{2}))L_{2}+
  \nonumber
\\
  &+&
  p^{2}(k^{2}+(kp))L_{3}+2((kp)^{2}-k^{2}p^{2})],
\label{off-shell-G1}
\\\nonumber
\\
  G_{2} &=& -\frac{m^{2}_{f}(kp)}{((kp)^{2}-k^{2}p^{2})^{2}}\times
  \nonumber
\\
  &\times&
  [(4m^{2}_{f}((kp)^{2}-k^{2}p^{2})+(k^{2}+p^{2})(2(kp)^{2}+k^{2}p^{2})+2(kp)((kp)^{2}+2k^{2}p^{2}))C_{0}+
  \nonumber
\\
  &+&
  (2(kp)^{2}+3(kp)k^{2}+k^{2}p^{2})L_{1}-(3(kp)(k^{2}+p^{2})+2(2(kp)^{2}+k^{2}p^{2}))L_{2}+
  \nonumber
\\
  &+&
  (2(kp)^{2}+3(kp)p^{2}+k^{2}p^{2})L_{3}+2((kp)^{2}-k^{2}p^{2})].
\label{off-shell-G2}
\end{eqnarray}

Now the off-shell amplitude can be calculated as
$M=T_{\mu\nu}k_{1\perp}^{\mu}k_{2\perp}^{\nu}/|{\bf
k}_{1\perp}||{\bf k}_{2\perp}|$ with $T_{\mu \nu}$ given by
Eq.(\ref{ampmain}) and $G_1$ and $G_2$ given by
Eqs.(\ref{off-shell-G1}) and (\ref{off-shell-G2}), respectively.

\section{Estimating the size of the off-shell effects}

In the so-called heavy fermion limit $m_{f}\rightarrow \infty$ we get
\begin{eqnarray}
G_{1}|_{m_{f}\rightarrow\infty}=\frac23, \quad
G_{2}|_{m_{f}\rightarrow\infty}=0 \label{limit-m-infty}
\end{eqnarray}
for gluons with {\it arbitrary} virtualities $k^2$ and $p^2$. The
first limit for $G_{1}$ coincides with the classical on-shell
result (\ref{series}).

It is easy to show also that the form factor $G_{1}$ in the limit
$k^{2},p^{2}\rightarrow 0$
\begin{eqnarray}
G_{1}|_{k^{2},p^{2}\rightarrow0}=\tau
 \left[1-\frac14(1-\tau)\ln\left(\frac{\sqrt{1-\tau}+1}{\sqrt{1-\tau}-1}\right)^{2}\right]\quad
 \textrm{for}\quad
 \tau=\frac{4m_{f}^{2}}{q^{2}}>1,
\label{G1-on-shell}
\end{eqnarray}
that coincides with $I_{1}$ (\ref{on-shellFF}) for the relevant
case of Higgs boson production from on-shell gluon fusion with
$q^{2}=(k+p)^2\simeq m_{h}^2,\; m_{f}=m_{top}$. In this limit the second
form factor $G_{2}$
\begin{eqnarray}
G_{2}|_{k^{2},p^{2}\rightarrow0} = -\tau\left[5-2\sqrt{1-\tau}
\ln\left(\frac{\sqrt{1-\tau}+1}{\sqrt{1-\tau}-1}\right)+\frac14(1+\tau)
\ln\left(\frac{\sqrt{1-\tau}+1}{\sqrt{1-\tau}-1}\right)^{2}\right] \; .
\label{G2-on-shell}
\end{eqnarray}
Expansions for $\tau\gg 1$ and $p^2,\,k^2=0$ have the following
form
\begin{eqnarray}
 G_{1}=\frac23
 \left[1+\frac{7}{30}\frac{1}{\tau}+\frac{2}{21}\frac{1}{\tau^{2}}+O\left(\frac{1}{\tau^3}\right)\right],\qquad
 G_{2}=-\frac{1}{45}\frac{1}{\tau}-\frac{4}{315}\frac{1}{\tau^2}+O\left(\frac{1}{\tau^3}\right).
\label{seriesG1G2}
\end{eqnarray}
Let's write analogous expansions with taking into account the
non-zeroth gluon virtualities.

The infinitely heavy fermion limits (\ref{limit-m-infty}) do not
contain at all the dimensional quantities such as
$q^{2},\,p^{2},\,k^{2},$ so there is no difference in the order of
limits: at first $q^{2}\rightarrow 0$ and then
$k^{2},\,p^{2}\rightarrow 0$ or vice versa. Therefore it is more
convenient to work with the dimensionless parameters defined as
\begin{eqnarray}
\chi=\frac{q^2}{4m^{2}_{f}},\qquad
\xi=\frac{p^2}{4m^{2}_{f}}<0,\qquad \eta=\frac{k^2}{4m^{2}_{f}}<0.
\end{eqnarray}
Then the on-shell limit $p^{2},k^{2}\rightarrow 0$ is equivalent to
$\xi,\,\eta\rightarrow 0.$ The heavy quark limit
(\ref{limit-m-infty}) corresponds to $\chi,\,\xi,\,\eta\rightarrow
0$. We can now take into account the terms of first order in $\xi$
and $\eta$ in (\ref{seriesG1G2}). In the case of Higgs boson
production on average
$m_{h}^{2}\gg |\langle k^{2} \rangle|,|\langle p^{2} \rangle|$
\cite{Szczurek}
so we have to take into account the powers of $\chi$ higher than
the powers of $\xi,\,\eta.$ Expansions of $G_{1}$ and $G_{2}$ in
up to second order in $\chi$ and first order in $\xi$ and $\eta$ give
\begin{eqnarray}
 G_{1}(\chi,\xi,\eta)&=&\frac23
 \left[1+\frac{7}{30}\chi+\frac{2}{21}\chi^{2}+\frac{11}{30}(\xi+\eta)+
 O\left(\chi^{3},\,\xi^{2},\,\eta^{2},\,\chi\xi,\,\chi\eta,\,\xi\eta\right)\right],
\nonumber
\\
 G_{2}(\chi,\xi,\eta)&=&-\frac{1}{45}(\chi-\xi-\eta)-\frac{4}{315}\chi^{2}+
 O\left(\chi^{3},\,\xi^{2},\,\eta^{2},\,\chi\xi,\,\chi\eta,\,\xi\eta\right)
\; .
\label{seriesG1G2_xieta}
\end{eqnarray}
We have compared the exact
form factors (\ref{off-shell-G1}) and (\ref{off-shell-G2}) with
their expansion counterparts (\ref{seriesG1G2_xieta}) for
realistic parameters $m_{top}=0.17\; \textrm{TeV},\;m_{h}=0.15\;
\textrm{TeV}.$ In conclusion, we can use the form factor expansions
up to $|\xi|,\,|\eta|=0.3$ with a maximal error of less than $1\%$.

To make our calculations useful for the case of Higgs boson
production let us turn to the physical parameters relevant for
proton-proton collisions. Introducing
$p=x_{1}p_{1}+k_{1\perp},\;k=x_{2}p_{2}+k_{2\perp},$ where
$k_{1\perp},\,k_{2\perp}$ are space-like four-vectors associated with
the transverse momenta of gluons ${\bf k}_{1\perp}$ and ${\bf k}_{1\perp}$,
$p_{1},\,p_{2}$ are the hadron momenta, so $(p_{1}p_{2})=s/2$, and
neglecting the hadron mass $m_{p}\ll m_{h},\,m_{top},$ we see that
$p^{2}\simeq - {\bf k}_{1\perp}^{2}<0$ and $k^{2}\simeq - {\bf
k}_{2\perp}^{2}<0.$ The subject of our analysis is the following
normalized projection of the amplitude
$M=T_{\mu\nu}k_{1\perp}^{\mu}k_{2\perp}^{\nu}/|{\bf
k}_{1\perp}||{\bf k}_{2\perp}|,$ given by the formula
\begin{eqnarray}
M(g^{*}g^{*}&\rightarrow&
H)=-i\delta^{ab}\frac{\alpha_{s}}{4\pi}\frac{1}{v}\biggl[(m_h^2+{\bf
k}_{1\perp}^2+{\bf k}_{2\perp}^2+2|{\bf k}_{1\perp}||{\bf
k}_{2\perp}|\cos{\phi})\cos{\phi}\;G_1- \label{proj}
\\
&-&\frac{2(m_h^2+{\bf k}_{1\perp}^2+{\bf k}_{2\perp}^2+2|{\bf
k}_{1\perp}||{\bf k}_{2\perp}|\cos{\phi})^2 |{\bf
k}_{1\perp}||{\bf k}_{2\perp}|}{(m_h^2+{\bf k}_{1\perp}^2+{\bf
k}_{2\perp}^2)^{2}}\;G_2\biggr], \nonumber
\end{eqnarray}
where $G_1,\,G_2$ can be taken from (\ref{seriesG1G2_xieta})
for not too large $| {\bf k}_{1\perp} |$ and $| {\bf k}_{2\perp} |$,
$\phi$ is the azimuthal angle between gluon transverse momenta
${\bf k}_{1\perp}$ and ${\bf k}_{2\perp},$ the transverse momentum
of the produced Higgs boson is ${\bf p}_{\perp}={\bf
k}_{1\perp}+{\bf k}_{2\perp}$ and the virtual gluon polarization
tensor has been taken in the form \cite{Gribov_Levin,Lipatov}
\begin{eqnarray*}
\sum\epsilon^{\mu}\epsilon^{*\,\nu}=\frac{k_{\perp}^{\mu}k_{\perp}^{\nu}}{{\bf
k}_{\perp}^2} \; .
\end{eqnarray*}
%
For brevity, we shall denote the first term of the amplitude
by ${\cal M}_1$ and the second term by ${\cal M}_2$.
Neglecting the second term in (\ref{proj}) and taking the on-shell
value of $G_1$ one obtains
\begin{eqnarray}
M_1&=&-i\delta^{ab}\frac{\alpha_{s}}{4\pi}\frac{1}{v}
(m_h^2+{\bf k}_{1\perp}^2+{\bf k}_{2\perp}^2
+2|{\bf k}_{1\perp}||{\bf k}_{2\perp}|\cos{\phi})\cos{\phi}\;I_1,
\label{onshM}
\\
I_1&=&G_1|_{k^2,p^2\rightarrow 0}\simeq G_1^0. \nonumber
\end{eqnarray}
This coincides with the amplitude obtained in Ref.~\cite{Lipatov}
within a numerical factor.

Let us now quantify some of the off-shell effects discussed above.

\subsection{Effect on form factors}

\begin{figure}[!hp]
\begin{minipage}{0.49\textwidth}
\epsfxsize=\textwidth\epsfbox{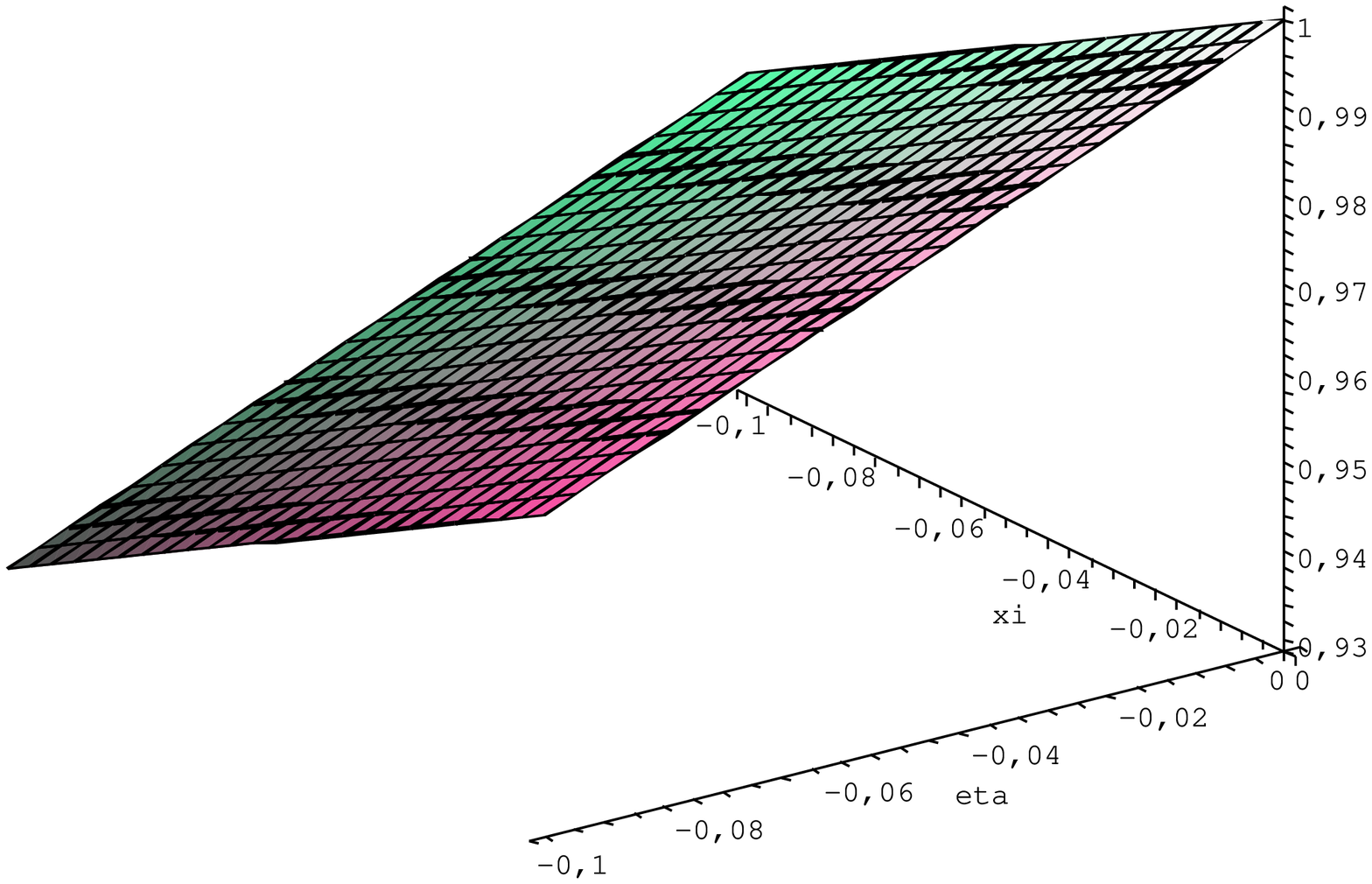}
\end{minipage}
\begin{minipage}{0.49\textwidth}
\epsfxsize=\textwidth \epsfbox{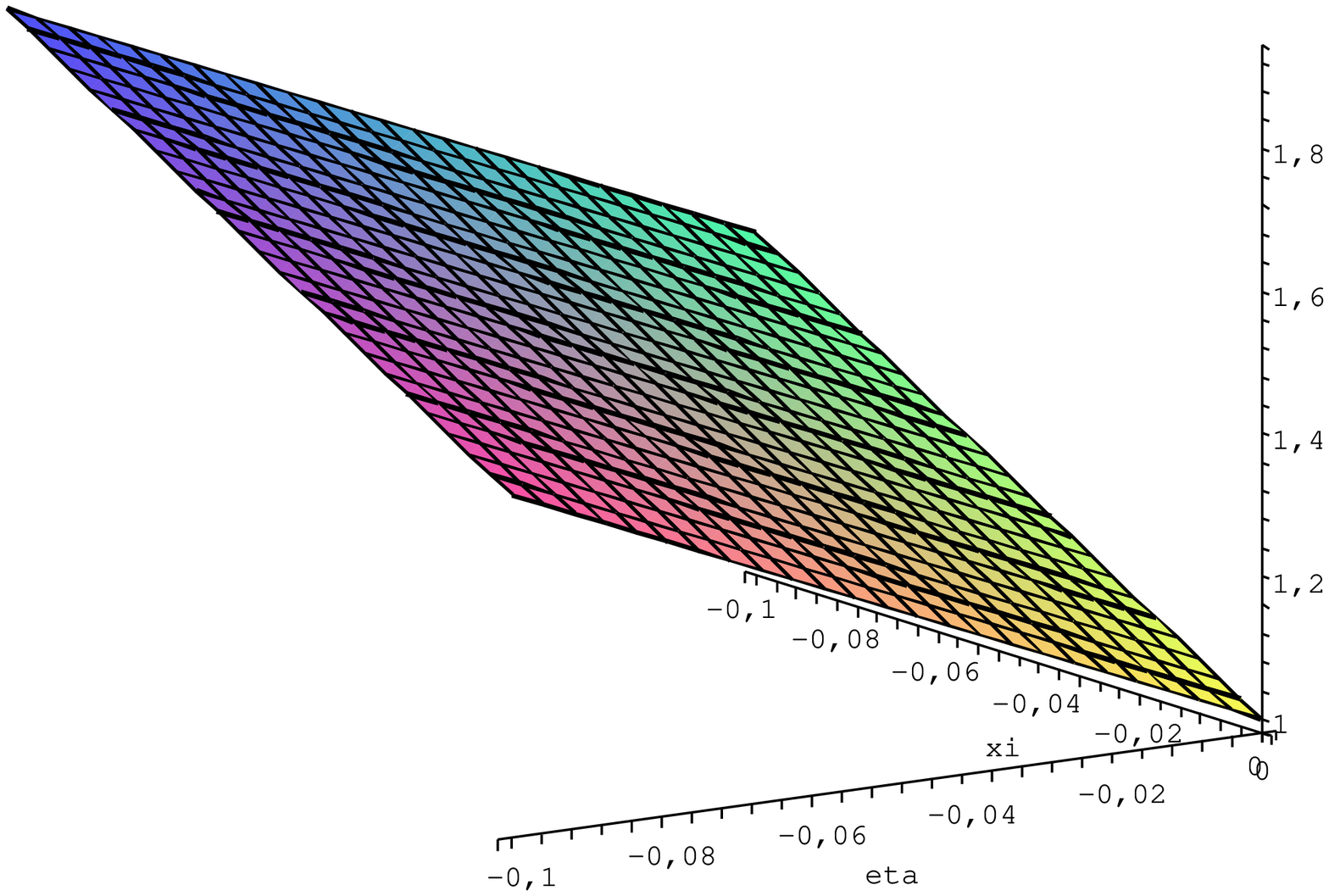}
\end{minipage}
\caption{\small Off-shell form factors $G_{1}$ (left) and $G_{2}$
(right) normalized to their on-shell values, as functions of
$\xi=-{\bf k}_{1\perp}^{2}/4m_{top}^2$ and $\eta=-{\bf
k}_{2\perp}^{2}/4m_{top}^2.$}
 \label{fig:fig2}
\end{figure}

Let us start with a simple case of form factors. In
Fig.~\ref{fig:fig2} we show the dependence of the two off-shell
form factors $G_1$ and $G_2$ on parameters $\xi$ and $\eta$. The
results are normalized to the on-shell values of the form factors
\begin{eqnarray}
G_1^0 &=& \frac{2}{3} \left[1 + \frac{7}{30} \chi + \frac{2}{21} \chi^2
\right] \; ,
\\ \nonumber
G_2^0 &=& -\frac{1}{45} \chi - \frac{4}{315} \chi^2 \; .
\label{on_shell_formfactors}
\end{eqnarray}
In this calculation \;$m_{top}$=0.17
\textrm{TeV},\;$m_{h}$=0.15\;\textrm{TeV}. We see that the first
form factor $G_{1}$ slightly drops, while the second form factor
$G_{2}$ grows with increasing $|\xi|$ and $|\eta|$. The first form
factor with taking into account of finite quark mass differs from
one obtained in the framework of effective approach
~\cite{Lipatov} by $5\,\%.$

\subsection{Effect on amplitude}

By averaging the amplitude squared $M^2$ over $\phi$ we obtain
\begin{eqnarray}
\langle M^2 \rangle_{\phi}
 &=& \frac{\alpha_{s}^2}{4\pi^2}\frac{1}{v^2}
\biggl[((m_h^2+{\bf k}_{1\perp}^2+{\bf k}_{2\perp}^2)^{2}
+2\,{\bf k}_{1\perp}^2{\bf k}_{2\perp}^2)\,G_1^2+ \nonumber
\\
 &+& \frac{8\,{\bf k}_{1\perp}^2{\bf k}_{2\perp}^2
[((m_h^2+{\bf k}_{1\perp}^2+{\bf k}_{2\perp}^2)^2
+6\,{\bf k}_{1\perp}^2{\bf k}_{2\perp}^2)^2
-32\,{\bf k}_{1\perp}^4{\bf k}_{2\perp}^4]}
{(m_h^2+{\bf k}_{1\perp}^2+{\bf k}_{2\perp}^2)^{4}}\;G_2^2-
\label{proj_average}
\\
&-&\frac{8\,{\bf k}_{1\perp}^2{\bf k}_{2\perp}^2
(3\,(m_h^2+{\bf k}_{1\perp}^2+{\bf k}_{2\perp}^2)^{2}
+2\,{\bf k}_{1\perp}^2{\bf k}_{2\perp}^2)}
{(m_h^2+{\bf k}_{1\perp}^2+{\bf k}_{2\perp}^2)^{2}}\;G_{1}G_{2}\biggr] \; .
 \nonumber
\end{eqnarray}
One can see that in the collinear limit ${\bf k}_{1\perp},\,{\bf
k}_{2\perp}\rightarrow 0$ the averaged square of matrix element
$\langle M^2 \rangle_{\phi}$ coincides with squared matrix element
in the covariant gauge
$T^{\mu\nu}T^{\mu'\nu'}g_{\mu\mu'}g_{\nu\nu'}$ multiplied by 2.

\begin{figure}[h]
 \centerline{\includegraphics[width=0.75\textwidth]{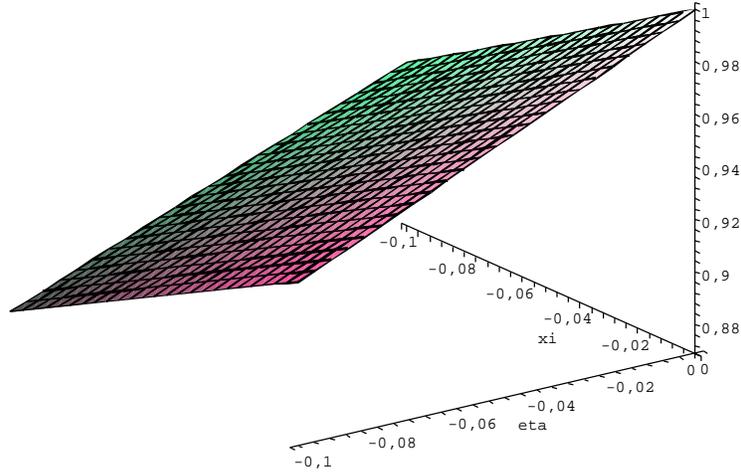}}
   \caption{\label{fig:fig3}
   \small Averaged square of off-shell matrix element $\langle M^2
\rangle_{\phi}$ normalized to its on-shell value $\langle M_0^2
\rangle_{\phi}$ as a function of $\xi=-{\bf
k}_{1\perp}^{2}/4m_{top}^2$ and $\eta=-{\bf
k}_{2\perp}^{2}/4m_{top}^2.$}
\end{figure}

Let's define the matrix element $M_0$ that is obtained from $M$ given by
(\ref{proj}) by substituting the off-shell form factors
by the on-shell ones $G_1^0$ and $G_2^0$
(see Eq.(\ref{on_shell_formfactors})).
\begin{eqnarray}
M_0&=&-i\delta^{ab}\frac{\alpha_{s}}{4\pi}\frac{1}{v}
\biggl[(m_h^2+{\bf k}_{1\perp}^2+{\bf k}_{2\perp}^2
+2|{\bf k}_{1\perp}||{\bf k}_{2\perp}|\cos{\phi})\cos{\phi}\;G_1^0-
\label{stmatrel}
\\
&-&\frac{2(m_h^2+{\bf k}_{1\perp}^2+{\bf k}_{2\perp}^2
+2|{\bf k}_{1\perp}||{\bf k}_{2\perp}|
\cos{\phi})^2 |{\bf k}_{1\perp}||{\bf k}_{2\perp}|}
{(m_h^2+{\bf k}_{1\perp}^2+{\bf k}_{2\perp}^2)^{2}}\;G_2^0\biggr], \nonumber
\end{eqnarray}
In Fig.~\ref{fig:fig3} we show $\langle M^2 \rangle_{\phi} /
\langle M_0^2 \rangle_{\phi}$ as a function of the reduced
parameters $\xi$ and $\eta$. One can see a drop of $\langle M^2
\rangle_{\phi}$ relative to $\langle M_0^2 \rangle_{\phi}$ due to
the drop of the form factor $G_1$.

Let us first estimate the off-shell effect. Following
Ref.~\cite{Szczurek} we take the values of transverse gluon
momenta in the interval $|{\bf k}_{1,2\perp}|\sim 5-50\;
\mathrm{GeV}$. Then the relative effect of replacing $G_1$ and
$G_2$ by their on-shell counterparts $G_1^0$ and $G_2^0$ is
\begin{eqnarray}
\frac{\langle M_0^2 \rangle_{\phi}-\langle M^2
\rangle_{\phi}}{\langle M_0^2
\rangle_{\phi}}\simeq\frac{(G_1^0)^2-G_1^2}{(G_1^0)^2}=0.0003 -
0.03 \; ,
\label{perc}
\end{eqnarray}
correspondingly for the lower and upper limits of gluon transverse
momenta.
A rather small drop of the averaged squared matrix element can be
observed, mainly due to the drop of the first form factor $G_1$.
The effect is of the order of $1\%$ or less at typical gluon transverse
momenta. Therefore the effect of non-zeroth gluon
virtualities in the form factors on the averaged squared
matrix element is rather small.

It seems interesting to study the behavior of the matrix
element in this part of the phase space where the first term is small.
Actually, when $\cos\phi=0$
the contribution of the form factor $G_1$ disappears from
(\ref{proj}), and we have the following simple expression for the
amplitude
\begin{eqnarray}
M|_{\phi=\pi/2}=i\delta^{ab}\frac{\alpha_{s}}{2\pi}\frac{1}{v}|{\bf
k}_{1\perp}||{\bf k}_{2\perp}| \;G_2, \label{offshell_part}
\end{eqnarray}
that is determined exclusively by the second form factor $G_2$.
As a result there are significant consequences of the non-zeroth
virtualities on the angular distribution around $\phi = \pi/2$.
Then $M^2|_{\phi=\pi/2}$ grows with increasing $|\xi|$ and $|\eta|$.
The relative growth is determined completely by the
growth of the form factor $G_{2}$
\begin{eqnarray}
\frac{M^2|_{\phi=\pi/2}-M^2_0|_{\phi=\pi/2}}{M^2_0|_{\phi=\pi/2}}
\simeq\frac{{G_2}^2-{G_2^0}^2}{{G_2^0}^2}=0.004 - 0.44,
\label{perc1}
\end{eqnarray}
at $|{\bf k}_{1,2\perp}|\sim 5-50\; \mathrm{GeV}$ respectively.
Thus taking into account the non-zeroth virtualities could
considerably change the angular distribution
at $\phi \approx \pi/2$. If this can be observed in reality
it will be discussed in the next section.

\subsection{Effect on inclusive cross section}

Let us try to evaluate how big can be the observable effect.
This requires a convolution of the off-shell subprocess cross section
with realistic unintegrated gluon distributions.
The inclusive cross section for the production of the Higgs boson
in hadron-hadron collisions can be written as
\begin{eqnarray}
\sigma = \int \frac{dx_1}{x_1} \frac{dx_2}{x_2}
\frac{d^{2} {\bf k}_{1\perp}}{\pi}
\frac{d^{2} {\bf k}_{2\perp}}{\pi}
\nonumber \\
\sigma_{off-shell}({\bf k}_{1\perp}, {\bf k}_{2\perp}) \;
{\cal A}(x_1,{\bf k}_{1\perp}^2,\mu_F^2) \;
{\cal A}(x_2,{\bf k}_{2\perp}^2,\mu_F^2) \; ,
\label{pp_tot_higgs}
\end{eqnarray}
where ${\cal A}(x,{\bf k}_{\perp}^2,\mu^2)$ are the unintegrated gluon
distributions in the colliding hadrons $h_1$ and $h_2$.

Now the distribution in azimuthal angle $\phi$
between ${\bf k}_{1\perp}$ and ${\bf k}_{2\perp}$ can be
calculated as:
\begin{eqnarray}
\frac{d \sigma}{d \phi} =
2 \pi
\int \frac{dx_1}{x_1} \frac{dx_2}{x_2}
\frac{1}{\pi^2}
k_{1\perp} d k_{1\perp}
k_{2\perp} d k_{2\perp}
\nonumber \\
\sigma_{off-shell}({\bf k}_{1\perp}, {\bf k}_{2\perp}) \;
{\cal A}(x_1,{\bf k}_{1\perp}^2,\mu^2) \;
{\cal A}(x_2,{\bf k}_{2\perp}^2,\mu^2) \; .
\label{dsig_dphi}
\end{eqnarray}
Inserting the off-shell matrix element squared to the formula for
the off-shell subprocess cross section we get
\begin{eqnarray}
\frac{d\sigma}{d\phi}&=&\frac{\alpha_s^2(\mu^2)}{256\,\pi^2}\,
\frac{(m_h^2+{\bf p}_{\perp}^2)}{v^2 x_1x_2sm_h^2}
\int
\biggl[\cos{\phi}\,G_1
-\frac{2(m_h^2+{\bf p}_{\perp}^2) |{\bf k}_{1\perp}||{\bf k}_{2\perp}|}
{(m_h^2+{\bf k}_{1\perp}^2+{\bf k}_{2\perp}^2)^{2}}
\;G_2\biggr]^2
 \label{diffcrosssec}
\\
&\times&{\cal A}(x_1,{\bf k}_{1\perp}^2,\mu^2)
        {\cal A}(x_2,{\bf k}_{2\perp}^2,\mu^2) \;
d\,{\bf k}_{1\perp}^2d\,{\bf k}_{2\perp}^2 d y_H,
\nonumber
\end{eqnarray}
where $y_H$ is the center-of-mass Higgs boson rapidity
and the longitudinal momentum fractions must be evaluated as
\begin{eqnarray*}
x_1=\sqrt{\frac{m_h^2+{\bf p}_{\perp}^2}{s}}\exp(y_H),\quad
x_2=\sqrt{\frac{m_h^2+{\bf p}_{\perp}^2}{s}}\exp(-y_H).
\end{eqnarray*}
Expression (\ref{diffcrosssec}), in the on-shell limit for form factors,
is consistent with the analogous formula obtained by Lipatov and Zotov
in Ref.~\cite{Lipatov}.

\begin{figure}[h]
 \centerline{\includegraphics[width=0.75\textwidth]{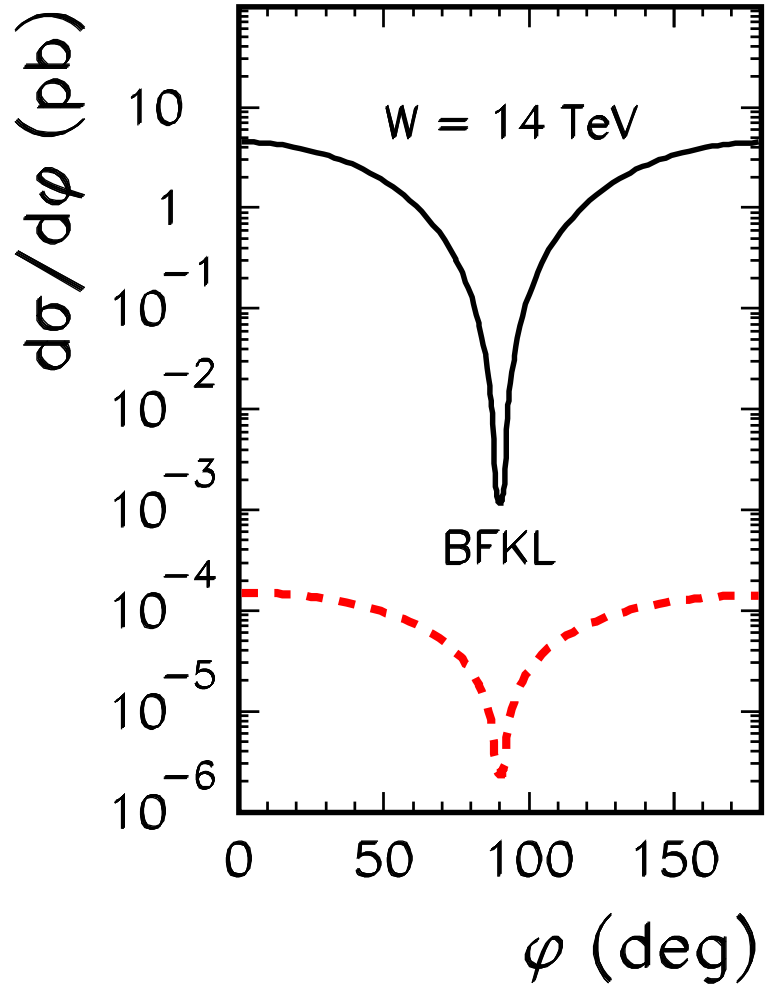}}
   \caption{\label{fig:fig4}
\small Azimuthal angle distribution of the cross section. In this
calculation the BFKL UGDF was used and $- 2  < y_H <  2$. The
solid line represents the calculation with the full amplitude,
whereas the dashed line is the modulus of the interference term.}
\end{figure}

The second term of the off-shell amplitude (\ref{proj}) is new and
was not discussed so far in the literature. For illustrating the
role of the second part of the amplitude in Fig.~\ref{fig:fig4} we
show the azimuthal angle distributions. In this calculation the
BFKL unintegrated gluon distributions were used for example (for
more details see e.g. \cite{Szczurek}). There is only a small
difference between the result obtained with the sum of both
amplitudes (${\cal M}_1 + {\cal M}_2$) and the result obtained
with the first term ( ${\cal M}_1$ ) only. The difference becomes
visible only close to $\phi = \pi/2$, i.e. when the first
amplitude vanishes. We show in the figure also the modulus of the
interference term. The latter is much smaller than the cross
section of the first term only, except of $\phi \approx \pi/2$.
Because this effect is negligible except of extremely close to
$\phi = \pi/2$, we see no easy way to identify the off-shell
effects.

\begin{figure}[h]
 \centerline{\includegraphics[width=0.75\textwidth]{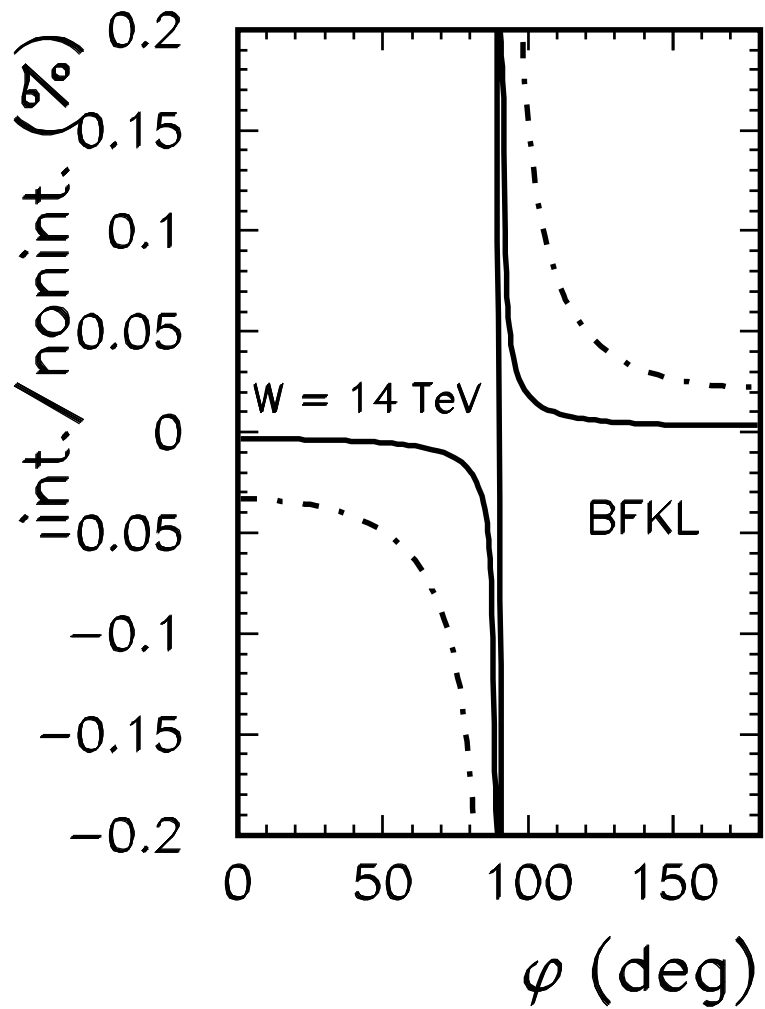}}
   \caption{\label{fig:fig5}
\small The asymmetry term normalized to the symmetric terms as a
function of $\phi$. In this calculation the BFKL UGDF was used and
$-2 < y_H < 2$. The solid curve corresponds to the inclusive case
while the dash-dotted curve is for the extra cut $|{\bf
p}_{\perp}| >$ 50 GeV. }
\end{figure}

From the theoretical point of view the exact off-shell matrix
element would break the familiar $\cos^2 \phi$-dependence of the
cross section. The second term of our exact matrix element leads
to an asymmetry around $\pi/2$:
\begin{equation}
\frac{d \sigma}{d \phi} \left( \frac{\pi}{2} - \phi_0 \right)
<
\frac{d \sigma}{d \phi} \left( \frac{\pi}{2} + \phi_0 \right)
\label{asymmetry}
\end{equation}
for $\phi_0 >$ 0. In Fig.~\ref{fig:fig5} we show the ratio of the
interference term to the sum of the two noninterference terms.
Here the off-shell effect is only at the $10^{-3}$ level. A bigger
effect is obtained at $\phi \approx \pi/2$. A change of sign at
$\phi = \pi/2$ is clearly visible. Putting an extra cut on $| {\bf
p}_{\perp} |$ could increase the relative asymmetry. In
Fig.~\ref{fig:fig5} we show the ratio obtained with such an extra
cut $| {\bf p}_{\perp} | >$ 50 GeV by the dash-dotted line.

\section{Conclusion}

In the present work we have analyzed the effect of the non-zeroth
virtualities of external gluons on the amplitude of
a scalar Higgs boson production.
An off-shell matrix element for this process was calculated.
We have found a new term in the amplitude compared to a recent
effective Lagrangian calculation.

A straightforward application of our analysis is in
the case of inclusive Higgs boson production in proton-proton collisions
at LHC. We have estimated that the relative drop of the averaged square
of the matrix element caused by replacing the on-shell form factors by
the off-shell ones is only about $1\%$ or less at relevant physical
parameters of the future experiments, so this effect
could be verified in the high-precision experiments only.
However, the effect of taking into account the non-zeroth
virtualities on the angular distribution at $\phi \approx \pi/2$ is
much more significant due to a quick growth of the second form factor
$G_2$ as a function of gluon transverse momenta.
The relative growth of the squared matrix element at $\phi=\pi/2$
is up to 45 \% for $|{\bf k}_{1,2\perp}|\sim 50\; \mathrm{GeV}$.

The observable effect was estimated in the framework of the
$k_{\perp}$-factorization approach by convoluting the squared
off-shell matrix element and the unintegrated gluon distributions.
The effect found is, however, extremely small and concentrated
around $\phi = \pi/2$. This will be extremely difficult to
identify in the future rather low-statistics experiments.

We have discussed deviations from the $\cos^2 \phi$-dependence of
the cross section. A small asymmetry around $\phi = \pi/2$, at the
$10^{-3}$ level, was found. Taking into account that the
$\phi$-dependence is not directly a measurable quantity the
predicted effect seems extremely difficult to be identified
experimentally.

\vspace {2cm}

We are grateful to A.E. Dorokhov, I.V. Anikin and J.
Ho\u{r}ej\u{s}\'i for useful discussions and comments. This work
was partially supported by the grant RFBR 03-02-16816
and by the grant of the Polish Ministry of Scientific Research
and Information Technology number 1 P03B 028 28.

\newpage



\newpage


\end{document}